\documentclass{optica-article}

\journal{opticajournal} 

\articletype{Research Article}

\usepackage{lineno}
\usepackage{soul}
\usepackage{xcolor}

\begin{document}

\title{Sub-Pixel Scale Structured Illumination for Lateral Resolution Enhancement of Non-Diffraction-Limited Flow Imaging}

\author{Hy Cao,\authormark{1,2} Abhishek Saha,\authormark{1,*} and Lisa V. Poulikakos\authormark{1,2,*}}

\address{\authormark{1}Department of Mechanical and Aerospace Engineering, UC San Diego, La Jolla, CA, USA\\
\authormark{2}Program of Materials Science and Engineering, UC San Diego, La Jolla, CA, USA\\}

\email{Corresponding authors: \authormark{*}asaha@ucsd.edu, lpoulikakos@ucsd.edu}


\begin{abstract*} 

In fluid flow imaging, intensity gradients are a good measure of spatial variations in scalar properties, which play an important role in controlling transport processes. However, current flow imaging techniques exhibit system-limited spatial resolutions, thus inhibiting the ability to accurately detect intensity gradients. To address this challenge, we present a method and system, inspired by Structured Illumination Microscopy (SIM), which can be implemented in dynamic flow imaging to enhance pixel resolution and, thereby, the estimation of scalar gradients. We utilize sub-pixel-scale patterned light matching the system pixel scale and multi-frame imaging that creates quasi-static images over four frames, with scalability for high-speed imaging. These multi-frame images are then processed using a bespoke recombination algorithm that produces a new image with twice the pixel resolution compared to the original images. The sub-pixel spatial-resolution enhancement capabilities are shown with static images and dynamic fluid flow, for which enhancement in the flow gradient is demonstrated. 
\end{abstract*}

\section{Introduction}

Image-based techniques are highly effective methods for quantitative measurement and evidence gathering in scientific research. While their accuracy closely depends on spatial resolution, any given imaging system has its spatial resolution constrained by either the diffraction of light or optical and hardware limitations~\cite{price_digital_2011}. The former is of concern, e.g., in biological imaging with high-magnification microscopes, where pixel sizes are well beyond the lower bound of the spatial resolution, resulting in over-sampled images with unresolvable detail\cite{pawleyHandbookBiologicalConfocal2006}. However, in most applications, e.g., consumer cell phones and scientific cameras, the camera hardware limits the spatial details captured in images. This limitation on the size and number of pixels in a sensor is a result of manufacturing and physical constraints, resulting in crosstalk and thus poor signal-to-noise ratios~\cite{hideshi_abe_device_2004}. This provides a hard limit on the spatial resolution of a camera sensor. 


Enhancing spatial resolution is critical for improved analysis and understanding of static and dynamic phenomena. Myriad techniques have been developed under the umbrella of \textit{super-resolution} (SR) to overcome inherent imaging limitations \cite{mockl_superresolved_2014, chenSuperresolutionStructuredIllumination2023, cnossenLocalizationMicroscopyDoubled2020, gustafssonDoublingLateralResolution2000, huangBreakingDiffractionBarrier2010, schermellehGuideSuperresolutionFluorescence2010, parkSuperresolutionImageReconstruction2003, tianSurveySuperresolutionImaging2011}. Pioneering super-resolution work by Eric Betzig, Stefan W. Hell, and William E. Moerner culminated with the 2014 Nobel Prize in Chemistry and continues to have a major impact \cite{mockl_superresolved_2014}. 
Standard SR methodologies feature a trade-off between spatial and temporal resolution, requiring multiple low-resolution (LR) images to reconstruct a high-resolution (HR) image. This enhancement arises from extractable information that is not inherently accessible from a single image but is obtainable through a set of images, where each image contains unique and definable information from changes in the illumination, subject, or optical train. SR techniques can typically be divided into diffraction-limited and system- or instrument-limited systems. 

Diffraction-limited techniques include those that utilize structured illumination (SI) \cite{rust_sub-diffraction-limit_2006, chenSuperresolutionStructuredIllumination2023, cnossenLocalizationMicroscopyDoubled2020, gustafssonDoublingLateralResolution2000, huangBreakingDiffractionBarrier2010, schermellehGuideSuperresolutionFluorescence2010} to generate Moiré fringes that fold high into low-order spatial frequencies, which can be shifted and extracted in Fourier space. 
For example, Structured Illumination Microscopy (SIM), utilizes patterned light as a Ronchi ruling to double the spatial resolution in a diffraction-limited microscope \cite{gustafssonDoublingLateralResolution2000}. Other diffraction limited SR methodologies encompass Stochastic Optical Reconstruction Microscopy (STORM), Total Internal Reflection Fluorescence Microscopy (TIRFM), and Confocal Laser Scanning Microscopy (CLSM) \cite{rust_sub-diffraction-limit_2006, fish_total_2009, huangBreakingDiffractionBarrier2010}.

In instrument-limited configurations, the smallest resolvable features pertain to the sensor's pixel density and optical magnification. 
From the pixel resolution to the diffraction limit, the spatial details and information of the object can be extracted by utilizing techniques such as sensor pixel-shift, subject micro-scanning, or source illumination variation  \cite{cnossenLocalizationMicroscopyDoubled2020, eladFastSuperresolutionReconstruction2001, fixlerPatternProjectionSubpixel2007, hardieHighresolutionImageReconstruction1998, luoPixelSuperresolutionUsing2016, parkSuperresolutionImageReconstruction2003, tianSurveySuperresolutionImaging2011, zhangSuperresolutionImagingInfrared2019}. 
These approaches typically define subject information with reference to each captured image to resolve additional physical information. Algorithmic processing then allows for resolution enhancement. 

Super-resolution approaches that employ recombination of multiple frames to obtain the reconstructed image also require efficient recombination algorithms. These include computationally heavy variants, alongside more recent machine learning and neural network approaches, enabled by advances in computational hardware \cite{vonchamierDemocratisingDeepLearning2021, eladFastSuperresolutionReconstruction2001, mullerOpensourceImageReconstruction2016, weigertContentawareImageRestoration2018, zhangSuperresolutionImagingInfrared2019}. Information extraction without a large set of images or even within a single image using structured light has also been demonstrated~\cite{kristenssonTwopulseStructuredIllumination2014, mishraThermometryAqueousSolutions2016}. Nonetheless, informational accuracy remains a function of the captured data with inherent limits to reconstructional image enhancements through computation. Thus, there is a need for physical systems that obtain images with higher resolution prior to algorithmic enhancements. 

In addition to enhancing image resolution beyond sensor capabilities or diffraction limits, most of the aforementioned techniques acquire multiple images of the same object with adjustments in background or lighting, thereby posing challenges for imaging dynamically moving objects. 
Motion is essential to the visualization and analysis of fluid flows, yet it brings about additional challenges. Image-based flow measurements often evaluate scalar gradients in, e.g., single-phase~\cite{prasad1990quantitative}, multi-phase\cite{saha2012breakup, saha2019kinematics}, reacting\cite{barlow2007laser, yang2016morphology}, and high-speed flows\cite{laurence2014schlieren}. The accuracy of the measured gradients is limited by image resolution. 
To address this challenge, structured illumination (henceforth denoted as SI) has been applied to fluid mechanics and similar dynamic phenomena via profilometry, to visualize surfaces or fluid-structure interactions \cite{florouSurfaceReconstructionThreedimensional2023, hoernerStructuredlightbasedSurfaceMeasuring2019}. Exploiting light outside of SI through evanescent waves has seen progress toward analysis of surfaces for interfacial flows \cite{yodaSuperResolutionImagingFluid2020}. 

Despite this, leveraging structured illumination for fluid mechanics to enhance information beyond physical or instrumentation constraints  remains largely unexplored. To date, structured illumination microscopy (SIM) has been applied to flows in microfluidic devices, generating phase shifts between a set of images, resulting in a two-fold improvement in spatial resolution \cite{luFlowbasedStructuredIllumination2013, spadaroResolutionConsiderationsStructured2023, spadaroStructuredIlluminationMicroscopy2020}. However, these studies face constraints to a specific small-scale flow in microchannels. In other studies, the Structured Laser Illumination Planar Imaging (SLIPI) method utilized structured light to suppress diffuse light, thereby increasing contrast in hollow-cone water spray imaging \cite{berrocalApplicationStructuredIllumination2008}. Further work on SLIPI opened avenues to increase information acquisition, such as droplet sizing and thermometry data \cite{mishraReliableLIFMie2014, mishraThermometryAqueousSolutions2016}. More recently, SI has been used to enhance the temporal resolution of Schlieren imaging by optically multiplexing a series of SI modulations into a single image, bypassing hardware limitations \cite{ekHighspeedVideographyTransparent2022}. However, several existing techniques address diffraction-limited systems rather than system- and instrumentation-limited constraints. Other imaging techniques that focus on system-limited spatial resolution are not suitable for dynamic subjects, particularly for macroscale flows with dimensions of millimeters or more.

To address these challenges, this work introduces a new structured illumination methodology that directly addresses system limitations to spatial resolution in flow imaging, by utilizing sub-pixel illumination patterns. While pixel sub-division has been applied to static subjects \cite{fixlerPatternProjectionSubpixel2007}, this work develops a methodology via linearly shifting SI patterns to enhance the lateral spatial resolution of consecutive images for dynamically moving subjects. We then apply a noise filtering algorithm, expanding on methodologies developed for infrared micro-scanning optical systems \cite{zhangSuperresolutionImagingInfrared2019}. We demonstrate this method by manufacturing a low-cost system that utilizes linearly driven gratings to enhance the spatial resolution of static and dynamic objects across a sensor by a factor of 2. The capabilities of the system are showcased through both static samples and, subsequently, dynamic tests with fluid flow, where the enhancement of the flow gradient is quantified.

\medskip

\section{Background and Experimental Setup}

In any given optical train, there exist two idealized lateral spatial resolution limits. A physical diffraction limit is determined by the optics and the resulting diffraction of light, while the camera pixel size and optical magnification determine a mechanical limit. The former (diffraction limit) is tied to the Airy disk, the smallest circle that an incoming point source of light can form through an aperture limited only by diffraction. The latter (mechanical limit) is applicable when the pixel size is well above the diffraction limit, resulting in a system limitation. The Nyquist–Shannon sampling criterion \cite{shannon_communication_1949} in signal processing states that the minimum sampling frequency to resolve a signal must be twice the frequency of the smallest desired signal. For imaging, this criterion dictates that the minimum spatial sampling frequency is twice that of the smallest desired features. In practice, it is common to use a factor of 2.3 times the minimum feature frequency to overcome noise arising from digital system imperfections~\cite{price_digital_2011}.

In diffraction-limited systems, where the sampled spatial frequency is well above the Nyquist–Shannon criterion, the spatial resolution limit is given by the minimum separation distance between two distinctly imaged Airy disks. The Abbe limit and the Rayleigh criterion define the minimum separation distance based on the Airy disks. While these formulas are functionally similar, the Abbe limit is based on the full-width half-maximum between two Airy disks. In contrast, the Rayleigh criterion is based on the distance from the center to the first minimum of the Airy disk. These values $d_{A}$ for the Abbe and $d_{R}$ for the Rayleigh criterion are defined by the numerical aperture ($\mathit{NA}$) and wavelength \(\lambda\) of light as shown in Eqs. \eqref{eq:abbe_limit} and \eqref{eq:rayleigh_criterion}, where $n$ is the refractive index and $\theta$ is the objective opening angle as 

\medskip
\begin{equation}
    \label{eq:abbe_limit}
    d_{A} = 
    \frac{0.5\lambda}{\mathit{NA}} 
    = \frac{0.5\lambda}{nsin(\theta)},
\end{equation}

\begin{equation}
    \label{eq:rayleigh_criterion}
    d_{R} = 
    \frac{0.61\lambda}{\mathit{NA}} = 
    \frac{0.61\lambda}{nsin(\theta)}.
\end{equation}
\medskip

We define the pixel pitch as the distance between two pixels on the sensor and the pixel size as the physical distance captured by a pixel. The relation between the two is simple and is defined in Eq. \eqref{eq:pixelSize} where optical magnification is the ratio between the size of the real subject and its projection on the image sensor plane. 

\medskip
\begin{equation}
    \label{eq:pixelSize}
    \mathit{Pixel\ Size = \frac{Pixel\ Pitch}{Optical\ Magnification}} 
\end{equation}
\medskip

For imaging setups where the pixel size is significantly larger than the diffraction limit, the spatial resolution is tied to the pixel size, and any spatial resolution enhancements are achieved through decreasing the pixel pitch or increasing the optical magnification. For most cameras, current sensor manufacturing has approached limits constrained by physics, and decreasing pixel pitch can no longer be achieved by trivial means. Conversely, increasing the optical magnification is possible, but it decreases the field of view and axial spatial resolution. Therefore, enhancement through super-resolution techniques is of significance, particularly for imaging setups for dynamic phenomena where sensors and optics are not easily changed. 

We focus on the enhancement of lateral spatial resolution in instrument-limited (i.e. non-diffraction-limited) systems, while imaging dynamic phenomena via structured illumination. Starting with any given pixel, we sub-divide this pixel at the imaging plane into an $x$-by-$x$ square grid, illuminate each subdivision of the pixel with a spot of light consecutively, and capture a series of images corresponding to each illuminated sub-division. This results in $x^2$ images, with intensity information that can be mapped spatially to the $x$-by-$x$ grid if the illumination position is previously known. For $x=2$, this allows us to halve the pixel scale and thus double the image resolution in both lateral directions for any given pixel, while the methodology can be scaled upwards until we approach the diffraction limit.

Expanding this across the entire field of view of the image creates a sub-pixel scale SI pattern and would allow for spatial resolution enhancement by a factor of $x$ for the entire image. This results in a two-dimensional (2D) periodic lattice. For each spot in the 2D lattice, the spot size or pitch at the imaging plane, $\mathit{D_{spt}}$, must be at most the size of a single sub-division of a pixel, defined as

\medskip
\begin{equation}
    \label{eq:spotPitch}
    \mathit{D_{spt} = \frac{Pixel\ Pitch}{x}}.
\end{equation}
\medskip

Each pixel subdivision is illuminated by shaping the spacing of the sub-pixel apertures into a hexagonal lattice and shifting linearly. Each discrete linear shift by distance $\mathit{D_{spt}}$ illuminates different positions as the pattern repeats itself and the starting position is re-illuminated. A continuous series of images containing differences in illumination can be captured by synchronizing a camera to capture images when the sub-pixel lattice is at appropriate positions (based on reference markers). The resulting grating must be designed to match the pixel scale of a given optical setup and any changes to the pixel scale in a system will dictate a change to the grating. 
\begin{figure}[ht]
    \centering\includegraphics[width=12cm]{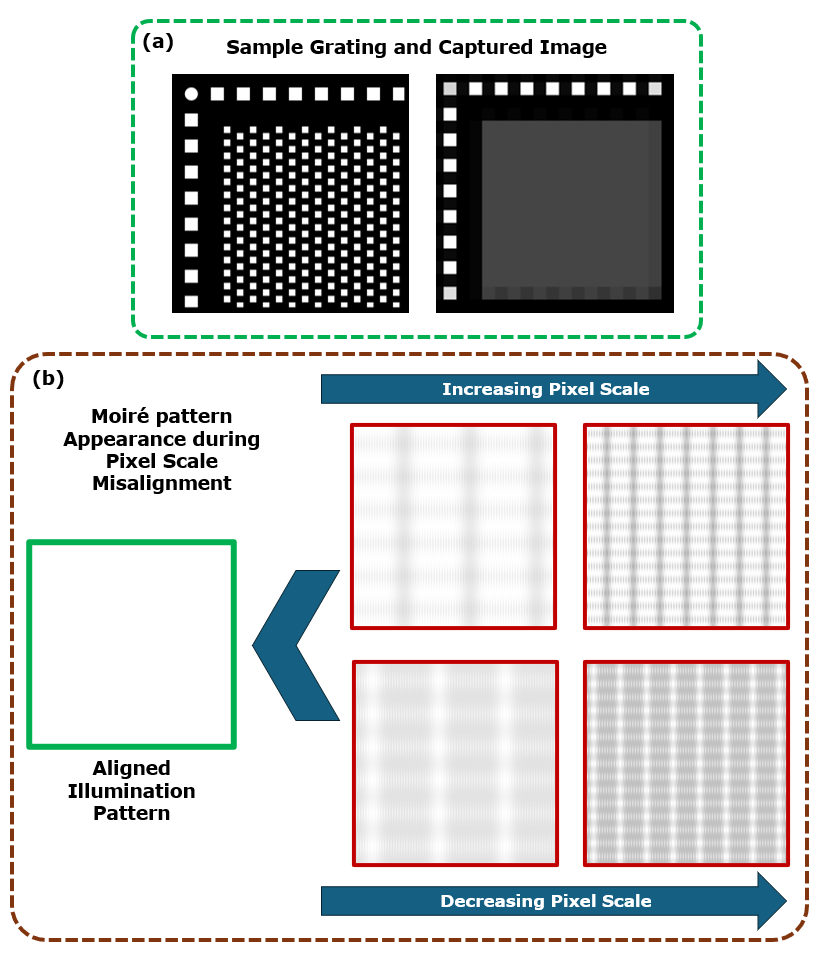}
    \caption{Grating example and visualization of misalignment during setup. (a) Sample grating with the expected captured image in the camera showing no visible sub-pixel pattern. (b) Simulation of the camera image arising in the case of an aligned illumination pattern (left) and Moiré patterns occurring when the system pixel size does not match the grating (right).}
    \label{fig:alignment}
\end{figure}

To effectively image with a sub-pixel grating, aligning the sub-pixel pattern with the camera sensor becomes critical. Since the individual illumination spots cannot be detected, structures defined relative to the grating above the Nyquist–Shannon criterion must be placed to allow for spatial definition of the lighting. These structures also serve to align the rotation, tilt, and starting position of the SI pattern to the imaging plane and sensor. It is important to align the grating at both the start and end of the linear translation range to ensure the axis of motion is running perpendicular to the camera optical axis. While necessary for alignment, these markers result in some loss of field of view for the final image, as no resolution enhancement can be performed in these areas. The markers of the studied illumination pattern are composed of a line of squares with a circle at the corner to quantify the physical distance between the borders and each hexagonal lattice spot. This grating design has two sections: the large structure borders and the sub-pixel hexagonal lattice. Fig.\ref{fig:alignment}(a) shows a sample grating with an arbitrary sub-pixel size and lattice pitch. The resulting simulated image capture has a pixel pitch double that of the sub-pixel lattice. The left image shows the hexagonally spaced lattice with larger alignment markers bordering the top and left edges. The right image shows how the larger alignment markers are still viable for camera capture, but the sub-pixel scale light becomes indistinguishable. 

Interestingly, a Moiré effect occurs when the sub-pixel grating pattern is not perfectly scaled to the camera sensor, which can be utilized for system alignment. Fig.\ref{fig:alignment}(b) shows this effect as the pixel scale becomes larger or smaller than that of the grating, with the leftmost image having a flat intensity profile when the sub-pixel pattern is perfectly aligned. This is because we can treat both the pattern and the sensor as a set of 2D gratings. When these gratings perfectly match, their frequencies align, and transmitted light generates a flat intensity profile. As the pixel size of a system changes, a mismatch occurs, resulting in lower-order frequencies appearing as a Moiré pattern in the image. 

For fluid flow and other dynamic phenomena, a set of $n$ images where $n>x^2$ can be captured and each successive set of $x^2$ images can be combined to create a set of $n-(x^2-1)$ reconstructed high-resolution images. Specifically, the first reconstructed high-resolution (rHR) image contains captured low-resolution (cLR) images 1 through $1+(x^2-1)$, the second contains images 2 through $2+(x^2-1)$, and the final possible reconstructed high-resolution (rHR) image contains low-resolution (cLR) images $n-(x^2-1)$ through $n$. It is important that the maximum subject movement between any $x^2$ number of frames is smaller than the spatial resolution of the high-resolution (rHR) images defined as
\medskip
\begin{equation}
    \label{eq:subjectMoveRate}
    \mathit{U_{sub}^{max} \leq \frac{Pixel\ Size * Frame\ Rate}{ x^2 }}.
\end{equation}

This is recommended to ensure quasi-static images. Otherwise, temporal aliasing will occur and provide poor data. Finally, for a linearly driven grating shifting the SI pattern, the grating must be driven at a velocity defined as

\medskip
\begin{equation}
    \label{eq:gratingMoveRate}
    \mathit{U_{grt} = \frac{Pixel\ Size * Frame\ Rate}{ x }}.
\end{equation}
\medskip

\begin{figure}[ht]
    \centering\includegraphics[width=12cm]{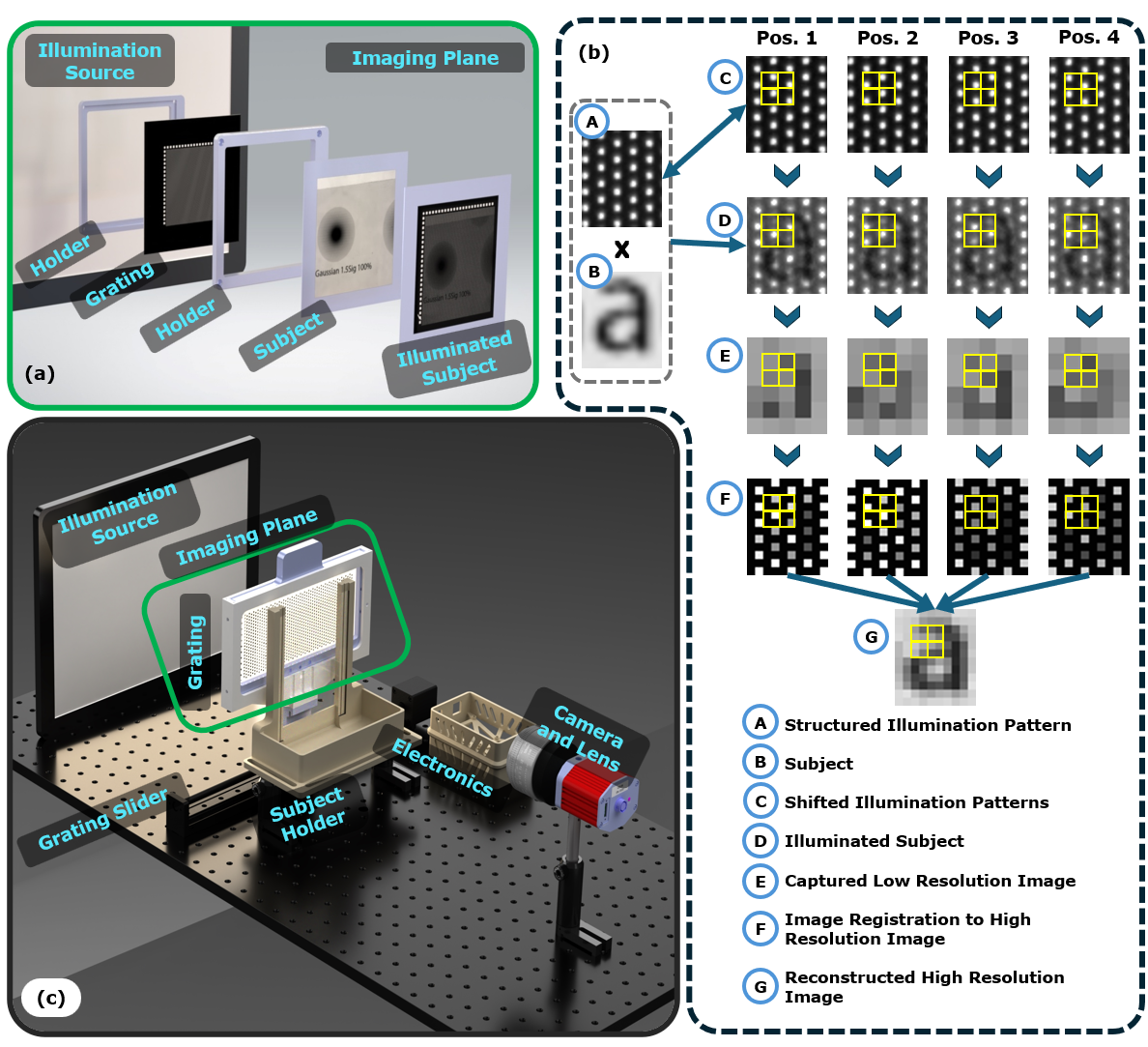}
    \caption{General overview of the system and method. (a) Illustration of the structured illumination pathway. (b) Schematic illustration of the sub-pixel structured illumination process in which a subject is illuminated with a sub-pixel pitch grating, captured with a camera, and combined to create a reconstructed high resolution image. (c) CAD model of the developed imaging system.}
    \label{fig:processFig}
\end{figure}

Figure \ref{fig:processFig} shows a general overview of the developed system and visualization of the imaging process. In Fig. \ref{fig:processFig}(a), the illumination pathway is shown as an exploded view with the illumination source, grating, grating holders, generic subject, and illuminated subject. Figure \ref{fig:processFig}(b) demonstrates the overall illumination chain for a pixel sub-division of 2. A and B are the SI pattern and subject, respectively. C shows the four shifts that A undergoes to generate the low-resolution (cLR) images. D is a visualization of the four SI pattern positions with their illumination onto subject B. E is the image captured by a camera, showing each individual frame without positional data. F is the expansion of each image after registering the position of each pixel relative to their known SI pattern position. Lastly, G is the final high-resolution (rHR) image. The set of 2-by-2 yellow boxes represents a set of four pixels across each step of the process. Figure \ref{fig:processFig}(c) shows a computer-aided-design (CAD) model of the finalized measurement system, including a motorized grating slider to generate linear translation of the illumination pattern for imaging of a simple jet flow. The area inscribed by the green rectangle represents the imaging plane visualized in Fig. \ref{fig:processFig}(a).

\section{Image Processing}

While simple image reconstruction is possible through quantification of the sub-pixel-scale illumination and subsequent superimposition of pixels from a low-resolution (cLR) image onto an high-resolution (rHR) grid, imprecision in the SI pattern alignment adds an additional source of noise compared to standard imaging techniques. The application of noise reduction becomes critical and is of particular importance in fluid mechanics. Therein, flow may have relatively low intensity variation between pixels, and quantification of gradients is fundamental to defining flow characteristics. We adopt a variation of the frequency domain phase-based projection onto convex sets (FPPOCS) SR algorithm developed for infrared micro-scanning systems \cite{zhangSuperresolutionImagingInfrared2019}. This algorithm relies on differences in the representation of real-space information in the Fourier space amplitude and phase spectrum \cite{oppenheimImportancePhaseSignals1981}. Textures and structures in real space are represented in the phase spectrum, where they are utilized to separate distinct features \cite{kovesiImageFeaturesPhase1999, zhangSuperresolutionImagingInfrared2019}. Note that the noise resulting from the SI pattern moves at a faster rate than the flow and has consistent features between frames. Thus, differences in the phase spectrum can be utilized to filter noise derived from the SI pattern.

\begin{figure}[ht]
    \centering\includegraphics[width=12cm]{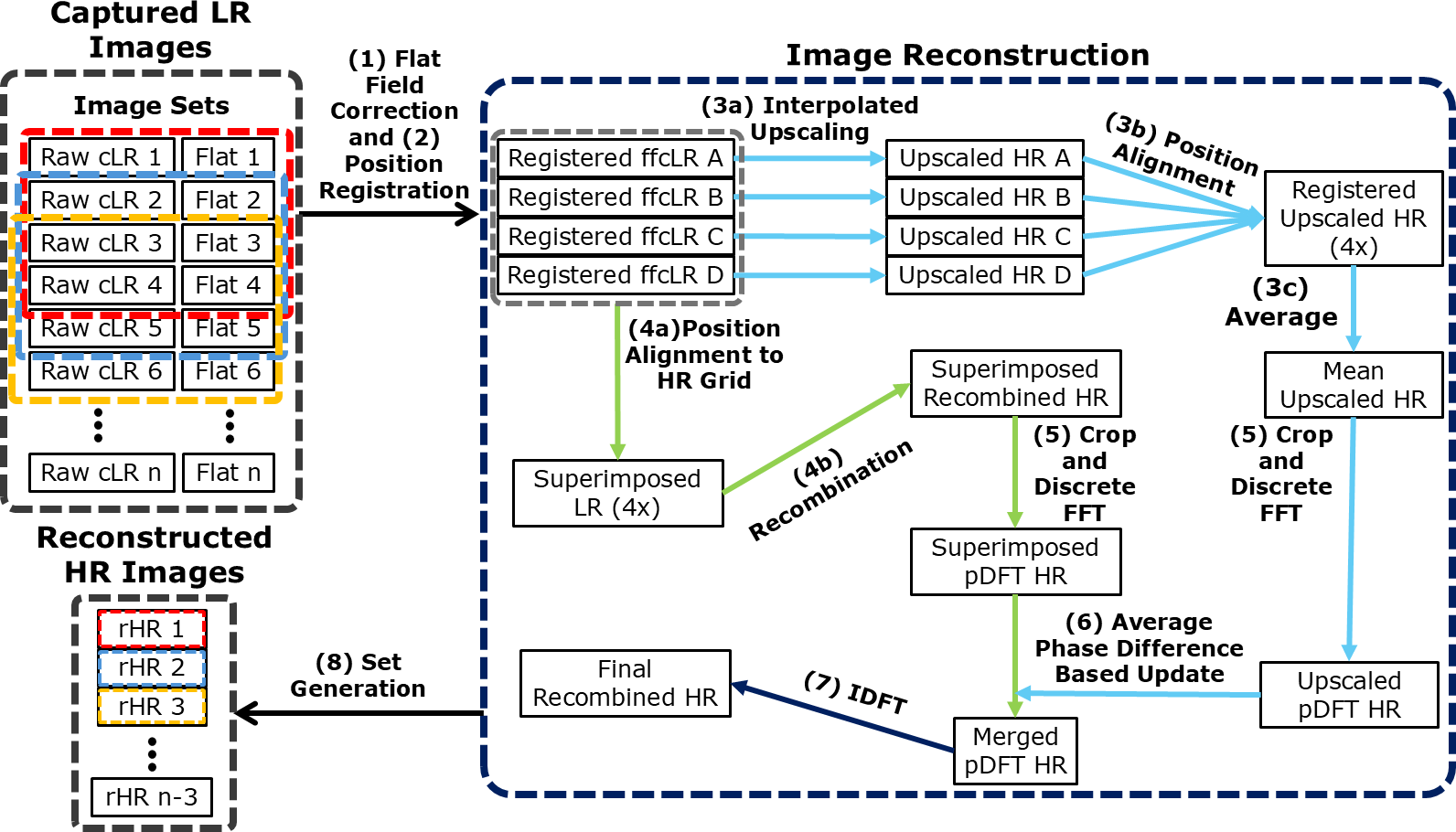}
    \caption{Workflow of the image reconstruction algorithm for a continuous set of n images. For each set of 4 images, an initial (1) flat field correction and (2) position registration are applied and calculated. (3a) Interpolated upscaling, (3b) position alignment, and (3c) average are performed on the 4 images to generate an upscaled HR image. (4a) Position alignment to HR grid and (4b) recombination are performed on the 4 images to generate a superimposed recombined HR image. (5) Crop and discrete fast Fourier transform (FFT) are applied on both the superimposed and upscaled HR images. Merging of the two images is achieved using an average phase difference-based update algorithm (6). An inverse discrete Fourier transform (IDFT) is applied (7), creating the resulting reconstruction HR image. Applying these steps to all consecutive sets of 4 images results in (8) set generation.}
    \label{fig:imageProcessFig}
\end{figure}

We define $x$ as the number of subdivisions a pixel undergoes in both lateral spatial axes, and summarize our image reconstruction and noise filtering technique for each $x^2$ set of images as described below. Moreover, Fig. \ref{fig:imageProcessFig} summarizes the image processing pipeline of our method with a sub-pixel division of $x=2$. We also utilize the same sub-division of $x=2$ in our proof-of-concept approach and setup. 

\begin{enumerate}
\item \textbf{Flat field correction (Fig. \ref{fig:imageProcessFig} (1), top left):} For each low-resolution (cLR) image, we capture a background image, in which imaging is done utilizing the system without a subject. The mechanical grating movement and camera parameters should be identical to the imaging done with the subject. This is considered as a "flat field". Similarly, dark current images for the low-resolution (cLR) image $D$ and flat field image $\mathit{FD}$ are captured . We then obtain a flat-field corrected LR image ($\mathit{ffcLR}$) by performing a standard correction for every image as defined in Eq. \eqref{eq:flatFieldCorr}. 

\medskip
\begin{equation}
    \label{eq:flatFieldCorr}
    \mathit{ffcLR = \frac{(cLR-D) * m}{F-FD}},
\end{equation}

\begin{equation}
    \label{eq:ffcMean}
    \mathit{m = mean(cLR-D)}.
\end{equation}
\medskip

\item \textbf{Position registration (Fig. \ref{fig:imageProcessFig} (2), top left):} As the grating design defines the distance of the border structures to the sub-pixel grating, the location of each sub-pixel lattice spot can be determined. The first image (of n total images) is utilized, and a corner of the sub-pixel lattice is selected as the origin to determine its precise coordinates within the pixel. The direction of the grating shift then determines the sub-pixel position in each consecutive frame with reference to the first.

\item \textbf{Interpolated HR image generation  (Fig. \ref{fig:imageProcessFig} (3a, 3b, 3c), top right):} For each ffcLR image, the image is resized by a factor of x through bi-cubic interpolation. Each image is spatially aligned, the average of all images is taken to obtain an estimated HR image $X_{est}$. 

\item \textbf{Superimposed HR image generation (Fig. \ref{fig:imageProcessFig} (4a, 4b), middle):} A HR array is generated that contains $x$ times the number of pixels along both lateral axes compared to the ffcLR image. For each ffcLR image, each pixel is then superimposed onto this HR array based on the known position and structure of the sub-pixel SI pattern. This is a simple geometric reconstruction with no noise filtering, containing only information from the base set of low-resolution (cLR) images that were defined as the preliminary HR image $X_0$. 

\item \textbf{Image Crop and discrete Fourier transform  (Fig. \ref{fig:imageProcessFig} (5), middle right):} For both $X_0$ and $X_{est}$, each image is first cropped to contain only the region of interest covered by the sub-pixel scale SI pattern. A padded zero discrete Fourier transformation (pDFT) is then applied to obtain $\tilde{X}_{0}$ and $\tilde{X}_{est}$ defined as:

\medskip
\begin{equation}
    \label{eq:phase1}
    \begin{cases}
        \tilde{X}_{0} = pDFT({X}_{0}) = A_{X_0}e^{i\varphi_{X_{0}}}\\
        \tilde{X}_{est} = pDFT({X}_{est}) = A_{X_{est}}e^{i\varphi_{X_{est}}}.\\
    \end{cases}
\end{equation}
\medskip

\item \textbf{Phase spectrum comparison and update  (Fig. \ref{fig:imageProcessFig} (6), bottom right):} Utilizing the frequency domain phase-based projection onto convex sets (FPPOCS) algorithm\cite{zhangSuperresolutionImagingInfrared2019}, the absolute difference of the phase angle between the two $\mathit{pDFT}$s for each wavenumber ${\Delta\varphi}$ is obtained from the phase spectrum. $m_{\Delta\varphi}$ is then the mean of all values in $\Delta\varphi$ as defined in Eqs. \eqref{eq:phase2} and \eqref{eq:phase3}. For values greater than $m$ in $\Delta\varphi$, the phase angle of that wavenumber is replaced in $\tilde{X}_{0}$ with the phase angle in $\tilde{X}_{est}$. This comparison is defined in Eq. \eqref{eq:phase4} and results in $\tilde{X}_{f}$:

\medskip

\begin{equation}
    \label{eq:phase2}
        {\Delta\varphi} = \mid \varphi_{X_{0}} - \varphi_{X_{est}}\mid ,
\end{equation}

\begin{equation}
    \label{eq:phase3}
        {m_{\Delta\varphi}} = \frac{1}{MN}\sum_{i=1}^N\sum_{i=1}^M{\Delta\varphi},
\end{equation}

\begin{equation}
    \label{eq:phase4}
    \tilde{X}_{f} =A_{X_0}
    \begin{cases}
        {e^{i\varphi_{X_{est}}}} \ \ \  \text{if} \ \ \ \Delta\varphi > m_{\Delta\varphi}\\
         {e^{i\varphi_{X_{0}}}} \ \ \ \ \text{otherwise}.\\
    \end{cases}
\end{equation}

\medskip

\item \textbf{Inverse Fourier Transform  (Fig. \ref{fig:imageProcessFig} (7) bottom middle):} Lastly, the inverse Fourier transform of $\tilde{X}_{f}$ is performed to obtain $X_{f}$, the reconstructed high-resolution image defined as:

\medskip
\begin{equation}
    \label{eq:phase5}
    {X}_{f} = iDFT(\tilde{X}_{f}).
\end{equation}
\medskip

\item \textbf{Set Generation (Fig.\ref{fig:imageProcessFig} (8) bottom left):} For each set of 4 captured low-resolution images in sequence, steps 3 through 7 are repeated to generate the full set of $n-3$ reconstructed HR images. 

\end{enumerate}

\newpage

\section{Experimental Methods}

Our approach focuses on facile, cost-effective manufacturing of a grating translation device that can be placed in a free-space optical train between an illumination source and a camera. This allows transparent subjects with variations in density, as is the case, e.g., in flow imaging, to be visualized by changes in illumination intensity transmission. As seen in Fig. \ref{fig:processFig}(c), we designed a setup in which a flat grating can be mounted in a 3D-printed holder attached to a 1-axis CNC stage driven by a stepper motor. Parts were all printed in polyethylene terephthalate glycol (PETG) utilizing a Creality Ender 3 V2, with part tolerances being plus or minus 0.1 $mm$. The 3D-printed grating holder enables efficient swapping with adjustment screws incorporated for rotational alignment around the imaging axis.   

Several design considerations arise due to mechanical limitations. The length of the grating slider determines the maximum continuous imaging time, as switching direction leads to acceleration and subsequent unusable frames. The length of the grating must also match the length of the slider and the precision of motion in the grating slider is critical. Lastly, utilization of broadband white light requires the subject holder to be placed as close as possible to the grating to minimize the spread of light.

We utilized a generic white LED flat panel as an illumination source, providing broadband uncollimated light. Structured illumination patterns were generated in MATLAB and printed onto a clear polyethylene terephthalate (PET) film using an ink-jet Epson ET-8550 to create gratings. This was placed into a removable 3D-printed holder attached to a FUYU FSL30 100 mm linear guide and rail, allowing for controlled, automated movement. Fine-screw adjustment of the grating holder relative to the imaging axes was integrated with springs applying tension to prevent shifting. 

Arduino DUO along with DM320T motor controller in a 3D-printed housing, was used for electronic control. The optical train contains a Nikon 50 mm f/1.8D adapted to a Thorlabs CS165CM camera with $3.45\mu m$ by $3.45\mu m$ pixels mounted onto a Thorlabs DTS50 translation stage for fine axial translation alignment. A generic LM393-based photoelectric sensor was integrated for stage homing and positioning. A simple Arduino script was programmed for motor control, and Thorlabs’ ThorCam software was used for image capture. 

To image transparent media and flow, a 3D-printed housing was fabricated, containing two glass plates creating a hollow channel. Static subjects, such as text on a transparent film, can be adhered to the glass plate for imaging. For flow, the housing was designed such that a tube could be sealed at the top of the channel. The bottom of the glass channel was placed such that a pool of water always remained above the exit, creating a sealed system. All of the air was pulled out of the system, resulting in a sealed vertical channel filled with water, where additional dyed water could be pushed from a syringe attached to a syringe pump, circumventing effects from gravity. The flow channel is 7 mm in thickness with a width of 86 mm and a height of 130 mm. An aluminum tube with an inner diameter of 5.63 mm was sealed at its top with silicone. The 60 ml syringe was attached to the aluminum tube using 6 mm tubing and pumped using a KDS 100 syringe pump. 

For our experiments, gratings with 105 $\mu m$ square holes spaced 352 $\mu m$ apart and 212 $\mu m$ square holes spaced 424 $\mu m$ apart were used. The former results in images with a low-resolution (cLR) pixel scale of 352 $\mu m$ and an high-resolution (rHR) scale of 176 $\mu m$. The latter results in images with a low-resolution (cLR) pixel scale of 424 $\mu m$ and an high-resolution (rHR) scale of 212 $\mu m$. During all imaging, the lens was set to an aperture ratio of f/4 with varied exposure times. Due to setup limitations and photo-printer-based grating limitations, the raw camera pixel scale and size were 8 times larger than the grating pitch sizes. Therefore, the resulting images were binned post-capture by a factor of 8. 

For lateral alignment, the location of the grating was determined visually via reference markers and adjusted during live image capture. Once the grating was level and its vertical positioning was set to ensure proper SI pattern positioning, the camera position was adjusted through the translation stage to ensure that the size of the sub-pixel SI matched the imaging plane. Lastly, a starting position was determined and programmed into the Arduino software. Further technical details regarding the setup can be found in the Supplemental Information section S1. After image capture, MATLAB was utilized to complete image processing. Binned images underwent the image processing steps described in the previous section.

\section{Results and Discussion}

\subsection{Imaging a static system: }
\begin{figure}[ht]
    \centering\includegraphics[width=12cm]{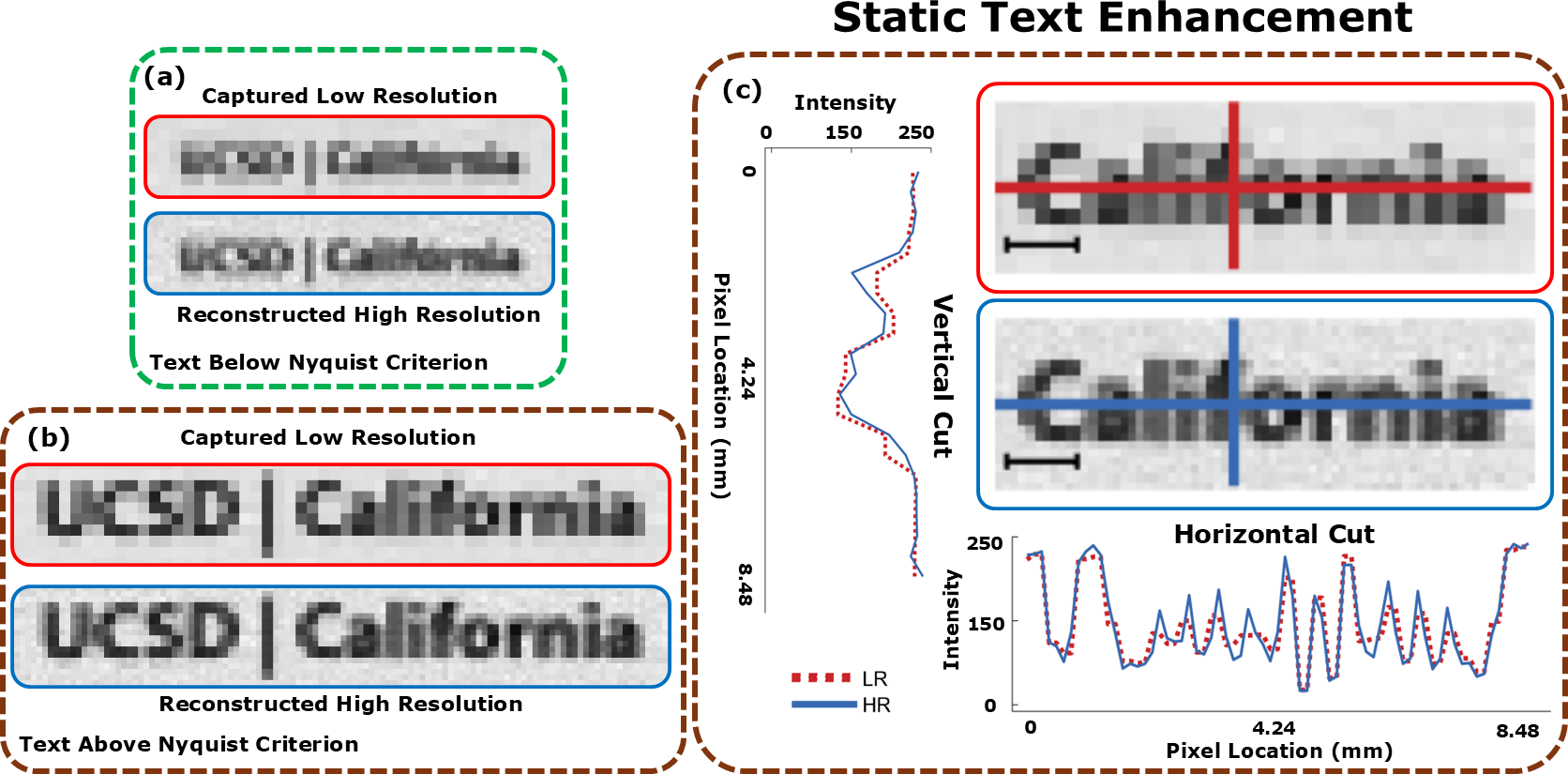}
    \caption{Example of spatial resolution enhancement for static text of the words 'UCSD' and 'California'. (a) Text enhancement at sampling rates below the Nyquist–Shannon criterion. (b) Text enhancement at sampling rates above the Nyquist–Shannon criterion. (c) Intensity plots for line-cuts of 'California' imaged at sampling rates above the Nyquist–Shannon criterion. Scale bars are 2.12 mm.}
    \label{fig:textResult}
\end{figure}

Before flow imaging, we verify the ability of our imaging system to enhance spatial resolution using static images of ink-jet printed texts onto a transparent PET film. We captured text containing the words ‘UCSD’ and ‘California’ with sampling rates below and above the Nyquist–Shannon criterion as shown in Fig. \ref{fig:textResult}a and b, respectively. The HR images are reconstructions from four images and were captured by static discrete shifts in the grating rather than continuous movement. Captured images were converted from the camera raw 16-bit color TIFF into a 16-bit monochromatic TIFF and binned by a factor of 8 prior to any image manipulation. After reconstruction, images were scaled for contrast and converted into 8-bit for more straightforward visualization. The resulting pixel scale of the LR image is 424 $\mu m$, and utilizing the more conservative Nyquist–Shannon criterion of 2.3 results in a spatial frequency of approximately 975 $\mu m$. The resulting reconstructed HR image has four times the pixel count, a pixel scale of 212 $\mu m$ and a resulting spatial frequency of approximately 488 $\mu m$. Both text sets are physically scaled to each other.  

In Fig. \ref{fig:textResult}(a), the low-resolution (cLR) image encircled in red contains largely illegible text. The high-resolution image, encircled in blue, shows a significant enhancement in legibility, with the capital letters being distinct and discernible. While the low-resolution (cLR) image in Fig. \ref{fig:textResult}(b) is mostly legible, some letters are missing features (e.g., 'f' and 'a'), or blended together (e.g., 'r' and 'n'). The resulting rHR image shows enhancement in features and contrast of the text with each letter being distinct. Fig. \ref{fig:textResult}(c) shows the word ‘California’ from (b) magnified for improved clarity (scale bar = 2 mm). Taking a line cut and plotting the intensity versus position across both the horizontal and vertical axes, we see enhancement in the peaks and valleys, resulting in the recovery of details that are not discernible in the low-resolution (cLR) image. In all cases, enhancement increases in the baseline noise floor seen in the background of the text. 

\subsection{Imaging a fluid system: }  
To demonstrate spatial resolution enhancement in fluid flow imaging, we created a small-scale, proof-of-concept experiment using a dyed water jet flowing into stagnant water. About 10 to 12 drops of water-soluble McCormick food dye were added per 120 $\mathit{ml}$ of deionized water for visualization. Due to its small quantity, the dye is expected to have negligible effects on the properties of water. In each experiment, using the syringe pump, the dyed liquid was pushed at a fixed volume flow rate through the tube with a 5.63 $\mathit{mm}$ diameter into the liquid channel holding stagnant clean water, creating a transient jet. The flow rate was varied between 60 and 160 $\mathit{ml/h}$. This results in the exit velocity of the jet ranging from 0.67 $\mathit{mm/s}$ to 1.785 $\mathit{mm/s}$ or a Reynolds number ($Re=UD/\nu$) range of $Re\in[3.9 - 10.5]$. 
Here, $U$ is the fluid velocity, $D$ is the diameter of the outlet, and $\nu$ is the kinetic viscosity of water at room temperature (22$^o$C). Due to its low $Re$, this simple jet flow retains a fairly laminar profile, developing into a mushroom-shaped tip, before expanding further (see Supplemental Information section S2). This flow feature is commonly observed when a liquid body (jet or droplet) enters a stagnant fluid environment \cite{saha2019kinematics, ChuaJet_2003, bhaskaran2025hydrodynamics}. 
The resulting profiles were captured from initial development through the jet traveling the length of the channel. We used a grating with 105 $\mu m$ square holes spaced 352 $\mu m$ apart. Imaging was performed at frame rates of 19.846, 39.692, and 40.05 fps with corresponding grating speeds of 3.5, 7, and 7.06 $\mathit{mm/s}$. Supporting Information tables S1 and S2 show in detail the flow and imaging parameters for each case analyzed. 

\begin{figure}[ht]
    \centering\includegraphics[width=10.5cm]{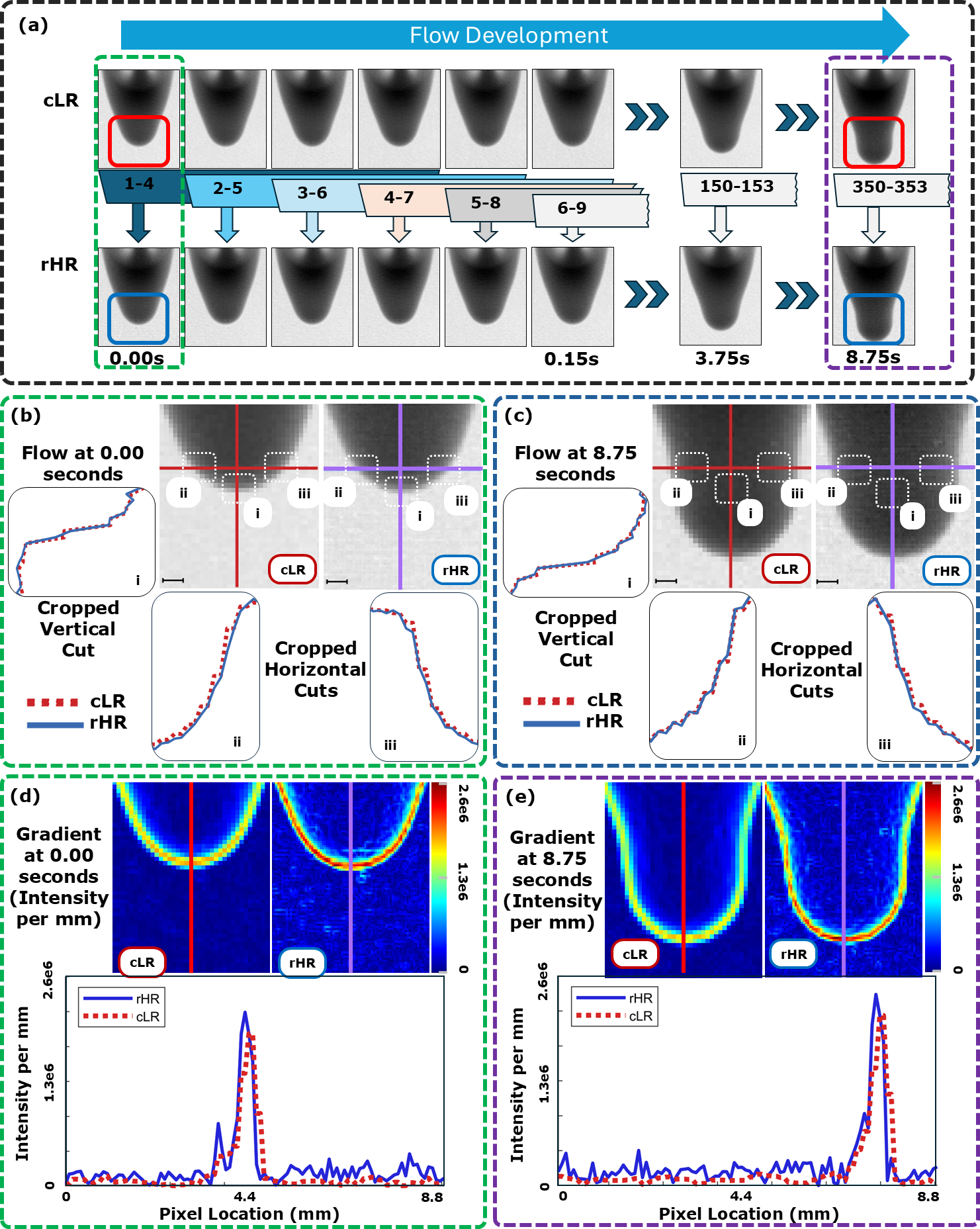}
    \caption{Flow imaging with an example mass flow rate of 160 $mm/h$ captured at 40 fps and enhanced with structured illumination. Flow enhancement is visualized through line cuts and gradient plots. (a) Captured low-resolution (cLR) and reconstructed high-resolution (rHR) images with labels denoting which frames are utilized to create a high-resolution (rHR) image for flow developing over 8.75 seconds. (b,c) Magnified image of the flow at 0 (part b) and 8.75 seconds (part c). Line-cuts in regions (i)-(iii) (shown in insets) exhibit an increase in the pixel resolution. Scale bars are 1.76 mm. (d,e) Flow gradient magnitudes for captured low-resolution and reconstructed high-resolution images at 0 (part d) and 8.75 seconds (part c). Gradients are shown as intensity per mm. Bottom panel: line cut of the center of each cropped gradient.}
    \label{fig:flowResult}
\end{figure}

Captured camera images were output as raw 16-bit color TIFFs. Since green dye was used, images were debayered, and the green image channels were extracted into a 16-bit monochromatic TIFF to obtain a higher signal-to-noise ratio. Images were then binned by a factor of 8 and separated into sets based on the changes in grating direction. After reconstruction, images were scaled for contrast. The resulting pixel scale of the low-resolution (cLR) image is 352 $\mu m$, where, utilizing the more conservative Nyquist–Shannon criterion, a spatial frequency of approximately 810 $\mu m$ is obtained. The resulting reconstructed HR image has four times the pixel count, a pixel scale of 176 $\mu m$ and a resulting spatial frequency of approximately 405 $\mu m$.

Figure \ref{fig:flowResult} shows the recombination of a set of images captured for the jet flow with a flow rate of 160 $\mathit{ml/h}$. Images were captured at 39.692 fps with the grating velocity at 7 $\mathit{mm/s}$. This is the maximum mass flow rate attainable for these imaging parameters before temporal aliasing occurs. Fig. \ref{fig:flowResult}(a) shows a timeline of the first captured frames along with snapshots of the flow as it develops. 8.75 seconds of flow progression are shown for this set, after which the grating device would switch directions. Both the low-resolution (cLR) and high-resolution (rHR) images are shown with blocks and text denoting which corresponding low-resolution (cLR) frames were used to create a given high-resolution (rHR) image. Frames 1-4 for the first high-resolution (rHR) image through to frame 350-353 for the last rHR image in the set are shown. 

Figure \ref{fig:flowResult}(b,c) focuses on a portion of the flow at the jet tip, where development is most prominent, at 0 seconds (part b) and 8.75 seconds (part c). For both sets, the magnified portions are marked by a rectangle in Fig. \ref{fig:flowResult}(a), with red denoting low-resolution (cLR) and blue denoting high-resolution (rHR) images. Insets in Figs. \ref{fig:flowResult}(b,c) show magnified line cuts of regions (i)-(iii) for the captured low-resolution (red, dashed) and reconstructed high-resolution (blue, solid) images, respectively, where the smoother slopes seen for rHR images demonstrate 2$\times$ the spatial resolution attainable for the low-resolution (cLR) image.

Fig. \ref{fig:flowResult}(d,e) quantifies flow gradient magnitudes before and after reconstruction at 0 (part d) and 8.75 seconds (part e). These gradients, calculated using the MATLAB gradient function, are visualized as a heat map. The color separates zones with the lowest gradient (equivalent to 0) marked by dark blue from zones with the highest gradient (approximately 260,000 intensity/mm) marked by dark red. The interface between the jet and the surrounding stagnant liquid contains high-gradient zones due to local shear effects. Furthermore, the high-resolution images display a higher density of information (resulting from smaller pixel sizes) and larger regions of higher gradient intensity, indicating an increase in available information. 
The bottom panels of Figs. \ref{fig:flowResult}(d,e) shows vertical line cuts for the captured low-resolution (red, dashed) and reconstructed high-resolution (blue solid) gradient images, indicating increased intensity for the rHR image, while also exhibiting an increase in the base noise floor. 

For quantitative measurements of jet features, we estimated the mixing-layer thickness marked by the thickness of the high-gradient zone from the line cuts. Measurement from the low-resolution (cLR) image yields a thickness of approximately 4.58 mm at both 0 and 8.75 seconds. In contrast, the enhanced high-resolution (rHR) images have a measured mixing-layer thickness of approximately 3.17 mm at both 0 and 8.75 seconds due to refinement of the captured data. In line with the static text enhancement (Fig. \ref{fig:textResult}), regions with little to no intensity change exhibit increased noise levels, as is particularly evident in the gradient background. The enhancement shown at 0 and 8.75 seconds was similar across all rHR images studied across all flow conditions. 

Analyzing flows captured for 80, 100, 120, and 160 ml per hour at 40.05 fps, we obtained an average flow tip length of 3.73 mm, 3.77 mm, 3.37 mm, and 3.33 mm, respectively (see Supplemental Information section S2 and Figs. S4 through S31). This aligns with the expectation that faster flows result in shorter mixing-layer thickness. It is worth noting that noise continues to play a role in these data points, and a smaller pixel scale would help refine the results further.    
    
\section{Discussion}
We conclude by summarizing this new method of improving image resolution and discussing the critical aspects of this technique. 
We demonstrated a structured illumination approach to achieving increased pixel resolution in an experimental configuration compatible with the in-situ visualization of moving objects, particularly for flows. Example cases of static text and dynamic fluid flow were demonstrated, where a 2$\times$ increase in spatial resolution was achieved. While the signal-to-noise ratio (SNR) of this technique is a challenge that the phase-spectrum algorithm seeks to mitigate, the noise floor seen in the background of all reconstructed images also shows an increase. This increase in noise is likely tied to several components of this technique, including imprecision in setup alignment, de-synchronization between motor movements and camera frame rate, and imperfections in the illumination from the grating. 

For setup alignment, the position of the patterned grid must be precisely aligned in both axes perpendicular to the camera axis. The pixel scale of the optical train must also match the grating as closely as possible. This alignment must be ensured at all positions of the linear motion of the grating, which is best done by checking the extremes of the motion. To circumvent mechanical errors (e.g., in motor movements) or de-synchronization, utilizing high-precision components with high repeatability, while ensuring a stable camera frame, is critical. Moreover, the linear motion of the grating generates a linearly shifting noise pattern akin to walking noise in astronomical imaging before and after recombination. While this is mitigated via flat-field correction, it proved difficult to remove entirely, as it may be tied to the accuracy of the grating alignment and movement. Lastly, imperfections in generating the structured illumination pattern also act as significant sources of noise. Our setup utilizes a simple ink on transparent film grating that works well for the studied pixel scales, yet imperfections from ink-jet printing result in variations in intensity between varying apertures of the structured illumination pattern, which is mitigated through flat-field correction.

Drawbacks to this technique are tied to the need for the SI pattern to match the imaging parameters. Any change to the optical train must also correlate to a change in the grating. This may be mitigated in the future by incorporating optics to resize and focus the SI pattern onto the imaging plane, or by changing how the SI pattern is generated. The upper bound of movement in the imaging phenomena is also inherently tied to the maximum camera frame rate, grating size, and grating speed. Lastly, since structures larger than the Nyquist–Shannon criterion are utilized for alignment, a reduction in field of view occurs. This reduction is typically a line 20 to 50 pixels in width across both the x and y axes.  

To conclude, through the utilization of a linearly driven hexagonal lattice SI pattern at a sub-pixel scale, we have demonstrated an enhancement of the fundamental spatial resolution for both static text and dynamic fluid flow. Our experimental setup obtains results that halve the pixel size and thus double the spatial resolution of the low-resolution (cLR) images. This was achieved through the development of a simple 3D-printed system utilizing photo-printed gratings and off-the-shelf electronics that can be easily manufactured. A phase-spectrum-based noise reduction technique was adapted and then applied to reduce the influence of grating movements. Lastly, we focus our future studies on obtaining better performance, laying the groundwork to bring this technique into high-speed imaging, and expanding on the variety of dynamic phenomena to be imaged. 

\section{Backmatter}

\begin{backmatter}
\bmsection{Funding}

H.C. and L.V.P gratefully acknowledge support from the Air Force Office of Scientific Research Young Investigator Program Grant (FA9550-23-1-0263). A.S. acknowledges partial support from the US National Science Foundation CAREER award (CBET-2145210). 
\bmsection{Acknowledgments}
The authors would like to thank Alexander Parrish, Zaid Haddadin, and Saaj Chattopadhyay for helpful discussions and Paula Kirya for assistance in spectral measurements. 
\bmsection{Disclosures}

\bmsection{Data Availability Statement}
A Data Availability Statement (DAS) will be required for all submissions beginning 1 March 2021. The DAS should be an unnumbered separate section titled ``Data Availability'' that
immediately follows the Disclosures section. See the \href{https://opg.optica.org/submit/review/data-availability-policy.cfm}{Data Availability Statement policy page} for more information.

\noindent Data availability statements are not required for preprint submissions.

\bmsection{Supplemental document}

See Supplement 1 for supporting content. 

\end{backmatter}
\bibliography{MyLibrary}

\begin{thebibliography}{10}
\newcommand{\enquote}[1]{``#1''}

\bibitem{price_digital_2011}
W.~G.~J. Jerome, \enquote{Digital {Image} {Capture} for {Confocal} {Microscopy},} in \emph{Basic {Confocal} {Microscopy},}  R.~L. Price and W.~G. Jerome, eds. (Springer New York, New York, NY, 2011), pp. 155--186.

\bibitem{pawleyHandbookBiologicalConfocal2006}
\enquote{Handbook {{Of Biological Confocal Microscopy}},}  (Springer US, Boston, MA, 2006), pp. 265--279, 3rd ed.

\bibitem{hideshi_abe_device_2004}
{Hideshi Abe}, \enquote{Device technologies for high quality and smaller pixel in {CCD} and {CMOS} image sensors,} in \emph{{IEDM} {Technical} {Digest}. {IEEE} {International} {Electron} {Devices} {Meeting}, 2004.},  (IEEE, San Francisco, CA, USA, 2004), pp. 989--992.

\bibitem{mockl_superresolved_2014}
L.~Möckl, D.~C. Lamb, and C.~Bräuchle, \enquote{Super‐resolved {Fluorescence} {Microscopy}: {Nobel} {Prize} in {Chemistry} 2014 for {Eric} {Betzig}, {Stefan} {Hell}, and {William} {E}. {Moerner},} {\protect\JournalTitle{Angewandte Chemie International Edition}} \textbf{53}, 13972--13977 (2014).

\bibitem{chenSuperresolutionStructuredIllumination2023}
X.~Chen, S.~Zhong, Y.~Hou, \emph{et~al.}, \enquote{Superresolution structured illumination microscopy reconstruction algorithms: A review,} {\protect\JournalTitle{Light: Science \& Applications}} \textbf{12}, 172 (2023).

\bibitem{cnossenLocalizationMicroscopyDoubled2020}
J.~Cnossen, T.~Hinsdale, R.~{\O}. Thorsen, \emph{et~al.}, \enquote{Localization microscopy at doubled precision with patterned illumination,} {\protect\JournalTitle{Nature Methods}} \textbf{17}, 59--63 (2020).

\bibitem{gustafssonDoublingLateralResolution2000}
M.~G.~L. Gustafsson, D.~A. Agard, and J.~W. Sedat, \enquote{Doubling the lateral resolution of wide-field fluorescence microscopy using structured illumination,} in \emph{{{BiOS}} 2000 {{The International Symposium}} on {{Biomedical Optics}},}  J.-A. Conchello, C.~J. Cogswell, A.~G. Tescher, and T.~Wilson, eds. (San Jose, CA, 2000), pp. 141--150.

\bibitem{huangBreakingDiffractionBarrier2010}
B.~Huang, H.~Babcock, and X.~Zhuang, \enquote{Breaking the {{Diffraction Barrier}}: {{Super-Resolution Imaging}} of {{Cells}},} {\protect\JournalTitle{Cell}} \textbf{143}, 1047--1058 (2010).

\bibitem{schermellehGuideSuperresolutionFluorescence2010}
L.~Schermelleh, R.~Heintzmann, and H.~Leonhardt, \enquote{A guide to super-resolution fluorescence microscopy,} {\protect\JournalTitle{Journal of Cell Biology}} \textbf{190}, 165--175 (2010).

\bibitem{parkSuperresolutionImageReconstruction2003}
S.~C. Park, M.~K. Park, and M.~G. Kang, \enquote{Super-resolution image reconstruction: A technical overview,} {\protect\JournalTitle{IEEE Signal Processing Magazine}} \textbf{20}, 21--36 (2003).

\bibitem{tianSurveySuperresolutionImaging2011}
J.~Tian and K.-K. Ma, \enquote{A survey on super-resolution imaging,} {\protect\JournalTitle{Signal, Image and Video Processing}} \textbf{5}, 329--342 (2011).

\bibitem{rust_sub-diffraction-limit_2006}
M.~J. Rust, M.~Bates, and X.~Zhuang, \enquote{Sub-diffraction-limit imaging by stochastic optical reconstruction microscopy ({STORM}),} {\protect\JournalTitle{Nature Methods}} \textbf{3}, 793--796 (2006).

\bibitem{fish_total_2009}
K.~N. Fish, \enquote{Total {Internal} {Reflection} {Fluorescence} ({TIRF}) {Microscopy},} {\protect\JournalTitle{Current Protocols in Cytometry}} \textbf{50} (2009).

\bibitem{eladFastSuperresolutionReconstruction2001}
M.~Elad and Y.~{Hel-Or}, \enquote{A fast super-resolution reconstruction algorithm for pure translational motion and common space-invariant blur,} {\protect\JournalTitle{IEEE Transactions on Image Processing}} \textbf{10}, 1187--1193 (2001).

\bibitem{fixlerPatternProjectionSubpixel2007}
D.~Fixler, J.~Garcia, Z.~Zalevsky, \emph{et~al.}, \enquote{Pattern projection for subpixel resolved imaging in microscopy,} {\protect\JournalTitle{Micron}} \textbf{38}, 115--120 (2007).

\bibitem{hardieHighresolutionImageReconstruction1998}
R.~C. Hardie, K.~J. Barnard, J.~G. Bognar, \emph{et~al.}, \enquote{High-resolution image reconstruction from a sequence of rotated and translated frames and its application to an infrared imaging system,} {\protect\JournalTitle{Optical Engineering}} \textbf{37}, 247--260 (1998).

\bibitem{luoPixelSuperresolutionUsing2016}
W.~Luo, Y.~Zhang, A.~Feizi, \emph{et~al.}, \enquote{Pixel super-resolution using wavelength scanning,} {\protect\JournalTitle{Light: Science \& Applications}} \textbf{5}, e16060--e16060 (2016).

\bibitem{zhangSuperresolutionImagingInfrared2019}
X.~F. Zhang, W.~Huang, M.~F. Xu, \emph{et~al.}, \enquote{Super-resolution imaging for infrared micro-scanning optical system,} {\protect\JournalTitle{Optics Express}} \textbf{27}, 7719 (2019).

\bibitem{vonchamierDemocratisingDeepLearning2021}
L.~{von Chamier}, R.~F. Laine, J.~Jukkala, \emph{et~al.}, \enquote{Democratising deep learning for microscopy with {{ZeroCostDL4Mic}},} {\protect\JournalTitle{Nature Communications}} \textbf{12}, 2276 (2021).

\bibitem{mullerOpensourceImageReconstruction2016}
M.~M{\"u}ller, V.~M{\"o}nkem{\"o}ller, S.~Hennig, \emph{et~al.}, \enquote{Open-source image reconstruction of super-resolution structured illumination microscopy data in {{ImageJ}},} {\protect\JournalTitle{Nature Communications}} \textbf{7}, 10980 (2016).

\bibitem{weigertContentawareImageRestoration2018}
M.~Weigert, U.~Schmidt, T.~Boothe, \emph{et~al.}, \enquote{Content-aware image restoration: Pushing the limits of fluorescence microscopy,} {\protect\JournalTitle{Nature Methods}} \textbf{15}, 1090--1097 (2018).

\bibitem{kristenssonTwopulseStructuredIllumination2014}
E.~Kristensson, E.~Berrocal, and M.~Ald{\'e}n, \enquote{Two-pulse structured illumination imaging,} {\protect\JournalTitle{Optics Letters}} \textbf{39}, 2584 (2014).

\bibitem{mishraThermometryAqueousSolutions2016}
Y.~N. Mishra, F.~Abou~Nada, S.~Polster, \emph{et~al.}, \enquote{Thermometry in aqueous solutions and sprays using two-color {{LIF}} and structured illumination,} {\protect\JournalTitle{Optics Express}} \textbf{24}, 4949 (2016).

\bibitem{prasad1990quantitative}
R.~R. Prasad and K.~Sreenivasan, \enquote{Quantitative three-dimensional imaging and the structure of passive scalar fields in fully turbulent flows,} {\protect\JournalTitle{Journal of Fluid Mechanics}} \textbf{216}, 1--34 (1990).

\bibitem{saha2012breakup}
A.~Saha, J.~D. Lee, S.~Basu, and R.~Kumar, \enquote{Breakup and coalescence characteristics of a hollow cone swirling spray,} {\protect\JournalTitle{Physics of fluids}} \textbf{24} (2012).

\bibitem{saha2019kinematics}
A.~Saha, Y.~Wei, X.~Tang, and C.~K. Law, \enquote{Kinematics of vortex ring generated by a drop upon impacting a liquid pool,} {\protect\JournalTitle{Journal of Fluid Mechanics}} \textbf{875}, 842--853 (2019).

\bibitem{barlow2007laser}
R.~S. Barlow, \enquote{Laser diagnostics and their interplay with computations to understand turbulent combustion,} {\protect\JournalTitle{Proceedings of the Combustion Institute}} \textbf{31}, 49--75 (2007).

\bibitem{yang2016morphology}
S.~Yang, A.~Saha, F.~Wu, and C.~K. Law, \enquote{Morphology and self-acceleration of expanding laminar flames with flame-front cellular instabilities,} {\protect\JournalTitle{Combustion and Flame}} \textbf{171}, 112--118 (2016).

\bibitem{laurence2014schlieren}
S.~Laurence, A.~Wagner, and K.~Hannemann, \enquote{Schlieren-based techniques for investigating instability development and transition in a hypersonic boundary layer,} {\protect\JournalTitle{Experiments in Fluids}} \textbf{55}, 1--17 (2014).

\bibitem{florouSurfaceReconstructionThreedimensional2023}
E.~I. Florou, C.~Fort, M.~A. Andr{\'e}, \emph{et~al.}, \enquote{Surface reconstruction in three-dimensional space using structured illumination,} {\protect\JournalTitle{Experiments in Fluids}} \textbf{64}, 70 (2023).

\bibitem{hoernerStructuredlightbasedSurfaceMeasuring2019}
S.~Hoerner and C.~Bonamy, \enquote{Structured-light-based surface measuring for application in fluid--structure interaction,} {\protect\JournalTitle{Experiments in Fluids}} \textbf{60}, 168 (2019).

\bibitem{yodaSuperResolutionImagingFluid2020}
M.~Yoda, \enquote{Super-{{Resolution Imaging}} in {{Fluid Mechanics Using New Illumination Approaches}},} {\protect\JournalTitle{Annual Review of Fluid Mechanics}} \textbf{52}, 369--393 (2020).

\bibitem{luFlowbasedStructuredIllumination2013}
C.-H. Lu, N.~C. P{\'e}gard, and J.~W. Fleischer, \enquote{Flow-based structured illumination,} {\protect\JournalTitle{Applied Physics Letters}} \textbf{102}, 161115 (2013).

\bibitem{spadaroResolutionConsiderationsStructured2023}
M.~Spadaro and M.~Yoda, \enquote{Resolution considerations for structured illumination microscale particle tracking velocimetry,} {\protect\JournalTitle{Experiments in Fluids}} \textbf{64}, 33 (2023).

\bibitem{spadaroStructuredIlluminationMicroscopy2020}
M.~Spadaro and M.~Yoda, \enquote{Structured illumination microscopy: A new way to improve the axial spatial resolution of microscale particle velocimetry,} {\protect\JournalTitle{Experiments in Fluids}} \textbf{61}, 127 (2020).

\bibitem{berrocalApplicationStructuredIllumination2008}
E.~Berrocal, E.~Kristensson, M.~Richter, \emph{et~al.}, \enquote{Application of structured illumination for multiple scattering suppression in planar laser imaging of dense sprays,} {\protect\JournalTitle{Optics Express}} \textbf{16}, 17870 (2008).

\bibitem{mishraReliableLIFMie2014}
Y.~N. Mishra, E.~Kristensson, and E.~Berrocal, \enquote{Reliable {{LIF}}/{{Mie}} droplet sizing in sprays using structured laser illumination planar imaging,} {\protect\JournalTitle{Optics Express}} \textbf{22}, 4480 (2014).

\bibitem{ekHighspeedVideographyTransparent2022}
S.~Ek, V.~Kornienko, A.~Roth, \emph{et~al.}, \enquote{High-speed videography of transparent media using illumination-based multiplexed schlieren,} {\protect\JournalTitle{Scientific Reports}} \textbf{12}, 19018 (2022).

\bibitem{shannon_communication_1949}
C.~Shannon, \enquote{Communication in the {Presence} of {Noise},} {\protect\JournalTitle{Proceedings of the IRE}} \textbf{37}, 10--21 (1949).

\bibitem{oppenheimImportancePhaseSignals1981}
A.~Oppenheim and J.~Lim, \enquote{The importance of phase in signals,} {\protect\JournalTitle{Proceedings of the IEEE}} \textbf{69}, 529--541 (1981).

\bibitem{kovesiImageFeaturesPhase1999}
P.~Kovesi, \enquote{Image {{Features}} from {{Phase Congruency}},} {\protect\JournalTitle{Videre: Journal of Computer Vision Research}} \textbf{1}, 1--26 (1999).

\bibitem{ChuaJet_2003}
L.~Chua, S.~Yu, and X.~Wang, \enquote{Flow visualization and measurements of a square jet with mixing tabs,} {\protect\JournalTitle{Experimental Thermal and Fluid Science}} \textbf{27}, 731--744 (2003).

\bibitem{bhaskaran2025hydrodynamics}
A.~M. Bhaskaran, A.~Paul, A.~Roy, \emph{et~al.}, \enquote{Hydrodynamics of liquid mushrooms,} {\protect\JournalTitle{Langmuir}}  (2025).

\end{thebibliography}

\end{document}